\documentclass[10pt,twocolumn,letterpaper]{article}

\usepackage{cvpr}
\usepackage{times}
\usepackage{epsfig}
\usepackage{graphicx}
\usepackage{amsmath}
\usepackage{amssymb}

\usepackage{caption}
\usepackage{subfigure}
\usepackage{color}
\usepackage{placeins}
\usepackage{float}
\usepackage{tabularx,colortbl}
\usepackage{times}
\usepackage{epstopdf}
\usepackage{mathrsfs}
\usepackage{diagbox}
\usepackage{multirow}
\usepackage{algorithm}
\usepackage{algorithmic}
\usepackage{galois}
\usepackage{url}

\cvprfinalcopy % *** Uncomment this line for the final submission

\def\cvprPaperID{19} % *** Enter the CVPR Paper ID here

\ifcvprfinal\fi
\begin{document}

%%%%%%%%% TITLE
\title{Deep Residual Learning for Image Compression}

\author{Zhengxue Cheng, Heming Sun, Masaru Takeuchi, Jiro Katto\\
Department of Computer Science and Communication Engineering, Waseda University, Tokyo, Japan\\
%Tokyo 169-8555, Japan.\\
{\tt\small \{zxcheng@asagi., hemingsun@aoni., masaru-t@aoni., katto@\}waseda.jp}
}

%% omit the following lines up until the closing ``}''.
%% Additional authors and addresses can be added with ``\and'',
%% just like the second author.
%% To save space, use either the email address or home page, not both
%\and

\maketitle
\vspace{-3mm}
%\thispagestyle{empty}

%%%%%%%%% ABSTRACT
\begin{abstract}
In this paper, we provide a detailed description on our approach designed for CVPR 2019 Workshop and Challenge on Learned Image Compression (CLIC). Our approach mainly consists of two proposals, i.e. deep residual learning for image compression and sub-pixel convolution as up-sampling operations. Experimental results have indicated that our approaches, \emph{Kattolab}, \emph{Kattolabv2} and \emph{KattolabSSIM}, achieve 0.972 in MS-SSIM at the rate constraint of 0.15bpp with moderate complexity during the validation phase.
%These finds are the basis of our submission to CVPR 2019 CLIC.
\end{abstract}

%%%%%%%%% BODY TEXT
\vspace{-3mm}
\section{Introduction}

Image compression has been an significant task in the field of signal processing for many decades to achieve efficient transmission and storage. Classical image compression standards, such as JPEG~\cite{IEEEexample:JPEG}, JPEG2000~\cite{IEEEexample:JPEG2000} and HEVC/H.265-intra~\cite{IEEEexample:HEVC}, usually rely on hand-crafted encoder/decoder (codec) block diagrams. Along with the fast development of new image formats and high-resolution mobile devices, existing image compression standards are not expected to be optimal and general compression solutions.

%They use fixed linear transform matrix, including Discrete Cosine Transform (DCT) and Discrete Wavelet Transform (DWT), together with the quantization and the entropy coder to reduce spatial redundancy for images. Usually, the standardization of a conventional codec has historically spanned several years. Along with the fast development of new image formats and high-resolution mobile devices, existing image compression standards are not expected to be optimal and general compression solutions.

Recently, we have seen a great surge of deep learning based image compression works. Some approaches use generative models to learn the distribution of images using adversarial training~\cite{IEEEexample:waveone, IEEEexample:MITgan, IEEEexample:Extreme}. They can achieve better subjective quality at extremely low bit rate. Some works use recurrent neural networks to compress the residual information recursively, such as~\cite{IEEEexample:Toderici01, IEEEexample:Toderici, IEEEexample:Nick}. These works are progressive coding, which can compress images at different quality levels at once. More approaches on relaxations of quantization and estimations of entropy model have been proposed in~\cite{IEEEexample:Theis, IEEEexample:Balle, IEEEexample:Balle2, IEEEexample:Balle3, IEEEexample:David, IEEEexample:softQuan, IEEEexample:conditional, IEEEexample:PCS, IEEEexample:CLIC, IEEEexample:HKPU}. Their ideas include using differentiable quantization approximation, or estimating the distribution for latent codes as entropy models, or de-correlating different channels for latent representation. Promising results have been achieved compared with classical image compression standards.

However, selecting a proper network structure is a daunting task for all types of machine learning tasks, including learned image compression. In this paper, we mainly discuss two issues. The first is about the kernel size. In classical image compression algorithms, filter sizes are quite important. Motivated from this, we conduct some experiments with different filter sizes to find larger kernel size contributes to better coding efficiency. Based on this observation, we propose to utilize deep residual learning to maintain the same receptive field with fewer parameters. This strategy not only reduces the model size, but also improves the performance greatly. On the other hand, the design of up-sampling operations at the decoder side is also significant to determine the reconstructed image quality and the type of artifacts. This issue has been widely discussed in super resolution tasks, and up-sampling layers can be implemented in various ways, such as interpolation, transposed convolution, sub-pixel convolution. We compare two popular up-sampling operations, i.e. transposed convolution and sub-pixel convolution to illustrate their performance.

%Motivated from the development of several super-resolution works, the up-sampling layers can be implemented in various ways, such as bicubic interpolation~\cite{IEEEexample:bicubicSR}, transposed convolution~\cite{IEEEexample:fastSR}, sub-pixel convolution\cite{IEEEexample:SubPixel}.

%Considering for fast end-to-end learning, we exclude bicubic interpolation and compare two popular up-sampling operations, i.e. transposed convolution and sub-pixel convolution to illustrate their performance.

%Results with CLIC validation dataset have been shown for our submitted entries.
%Experimental results have demonstrated the superior performance of our proposed approach.

In CLIC 2019, we submitted three entries including \emph{Kattolab}, \emph{Kattolabv2}, and \emph{KattolabSSIM} in the low rate track, to achieve 0.972 MS-SSIM with moderate complexity.

\section{Deep Residual Learning for Image Compression with Sub-Pixel Convolution}

\begin{figure}[tb]
\centering
\subfigure[Baseline-9]{
\label{Fig.anchor.1}
\includegraphics[height=85mm]{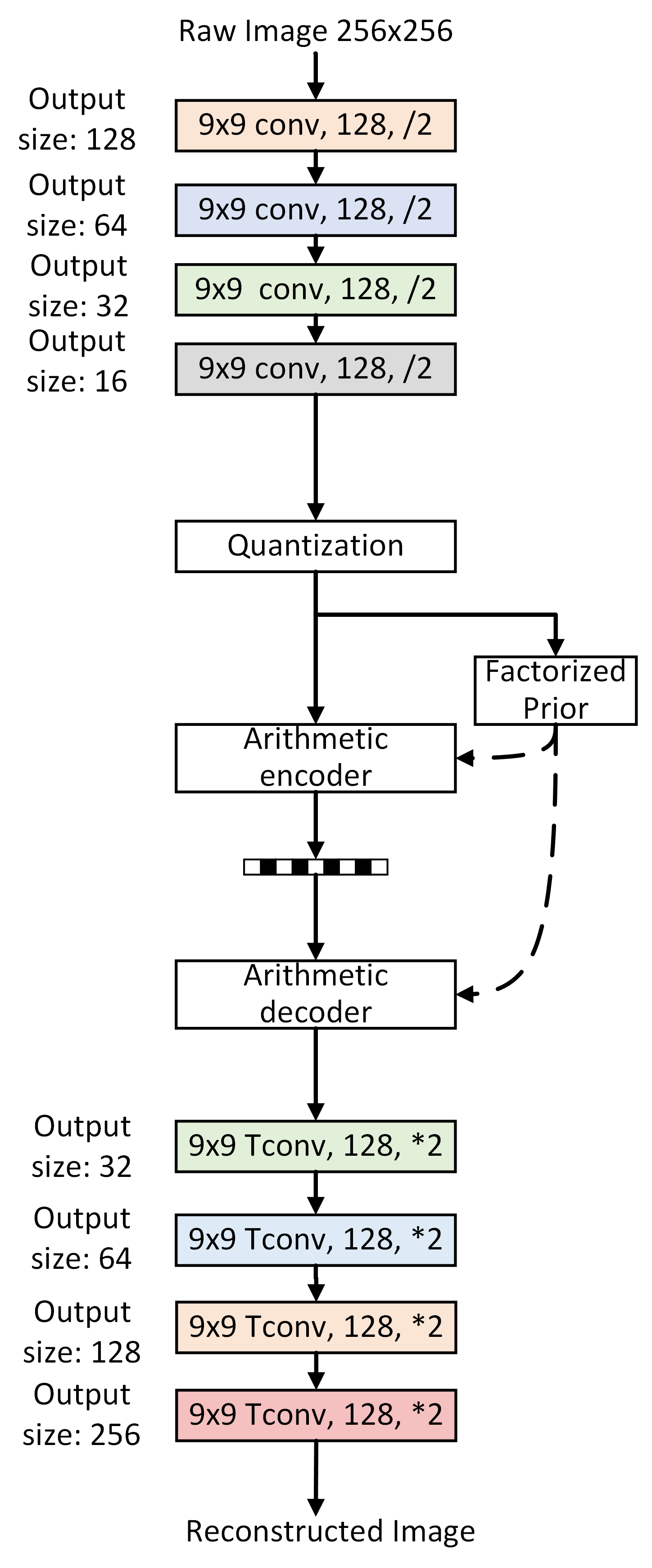}}
\subfigure[HyperPrior-9]{
\label{Fig.anchor.2}
\includegraphics[height=85mm]{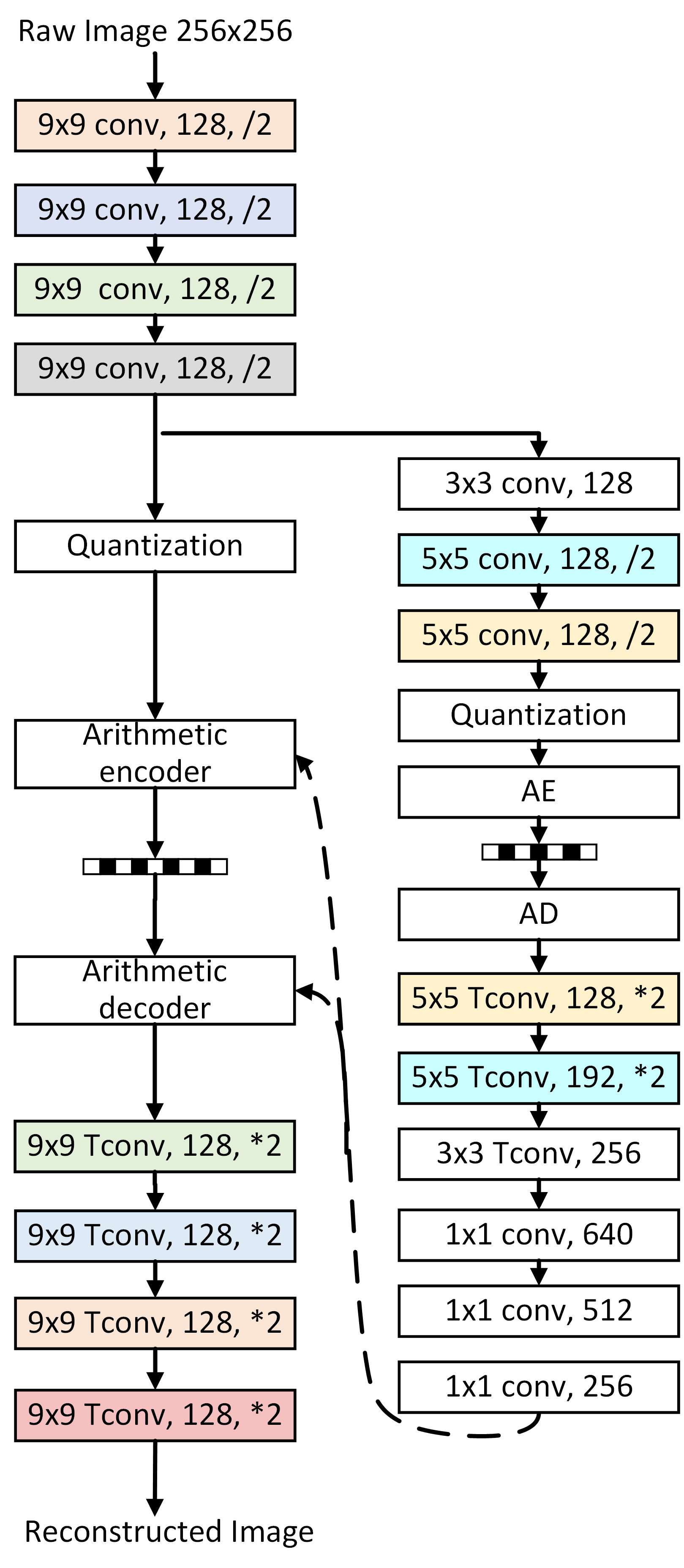}}
\caption{The network structure of anchors we used.}
\label{fig:anchor}
\end{figure}

The network architectures that we used as anchors are illustrated in Fig.~\ref{fig:anchor}. This architecture is referred from the work~\cite{IEEEexample:Balle2} and the work~\cite{IEEEexample:David}, which has achieved the state-of-the-art compression efficiency. The network consists of two autoencoders. The main autoencoder controls the rate-distortion optimization for image compression, and the loss function is formulated as
\begin{equation}
\begin{aligned}
\label{eq.loss}
J = \lambda d(\boldsymbol{x}, \hat{\boldsymbol{x}})+ R(\hat{\boldsymbol{y}})
%= \lambda d(\boldsymbol{x}, \hat{\boldsymbol{x}})+ \mathop{\mathbb{E}}_{\boldsymbol{x}\sim p_{x}}[-\log_{2}(p_{\hat{\boldsymbol{y}}}(\hat{\boldsymbol{y}}))]
\end{aligned}
\end{equation}
where $\lambda$ controls the tradeoff between the rate and distortion. The auxiliary autoencoder is used to encode the side information to model the distribution of compressed information. Gaussian scale mixture is used to estimate an image-dependent and adaptive entropy model, where scale parameters are conditioned on a hyperprior. Moreover, \cite{IEEEexample:David} proposed a joint autoregressive and hyperprior approach, denoted as \emph{Joint}. The only difference is to append a masked $5\times5$ convolution after quantization and to concatenate the output of auxiliary autoencoder and masked convolution together to learn the entropy model.

 %$d$ represents the distortion between the original images and reconstructed images , which can be mean square error (MSE) or MS-SSIM~\cite{IEEEexample:msssim} quality metrics.

%The overall learned image compression is shown in Fig.~\ref{fig:network}.
%This is the general architecture for image compression.
%
%We have also seen some masked CNN-based adaptive context works, including [David, BMVC, ICML2019], however, a sequential reconstruction is inevitable for decoding. We have also tested the performance.
%
%From the viewpoint of decoding time for CLIC challenge, we did not discuss masked CNN in this paper. But it can be incorporated into our proposed architecture easily.

%\begin{figure*}[tb]
%	\centerline{\psfig{figure=compare.PNG,width=180.0mm} }
%	\caption{The Network Architecture.}
%	\label{fig:network}
%\end{figure*}

\vspace{-2mm}
\subsection{From Small Kernel Size to Large Kernel Size}
\vspace{-2mm}

In classical image compression algorithms, transform filter sizes are quite important to improve the coding efficiency, especially for UHD videos. From the smallest transform size $4\times 4$, larger and larger transform size is gradually used into video coding algorithms. Specifically, up to $32\times32$ DCT coefficients have been incorporated into HEVC~\cite{IEEEexample:HEVC}. Large kernel size brings benefits on capturing the spatial correlation and semantic information. Motivated from this, we conduct some experiments using Kodak dataset~\cite{IEEEexample:kodak} with different filter sizes in the main and auxiliary autoencoders respectively to explore the effect of larger kernel size on coding efficiency. Table~\ref{Table.baseline} shows for the \emph{Baseline} architecture, along with the increasing of kernel sizes, the rate-distortion performance are becoming better. Table~\ref{Table.hyper} demonstrate similar results for \emph{HyperPrior} architectures. Table~\ref{Table.hypersub} shows large kernel in the auxiliary autoencoder cannot bring any benefits on RD performance and even gets worse, because the compressed codes $y$ has small size, so $5\times5$ are large enough. Too many learnable parameters instead increase the difficulty to learn. It is worth noting for \emph{Joint} architecture~\cite{IEEEexample:David}, a sequential decoding is inevitable, which is extremely time-consuming when the test image becomes larger. Therefore, we exclude the masked convolution in this challenge, but keep the $1\times1$ conv as they are, for \emph{HyperPrior} architecture. An ablation on the effect of $1\times1$ conv will be conducted in the future.

%Motivated from this, large kernel size should also be applied in learned image compression to learn the semantic information.

\begin{table}[tb]
\footnotesize
\begin{center}
\caption{The effect of kernel size on \emph{Baseline} on Kodak, optimized by MSE with $\lambda=0.015$.}
\label{Table.baseline}
\begin{tabular}{l|l|l|l}
 \hline
 \textbf{Method}    &\textbf{PSNR}& \textbf{MS-SSIM}  & \textbf{Rate} \\%&\textbf{Para}
 \hline
 \emph{Baseline-3}        &32.160   &0.9742 &0.671 \\ %&997379
 \hline
 \emph{Baseline-5}      &32.859   &0.9766 &0.641 \\%&2582531
 \hline
 \emph{Baseline-9}       &32.911   &0.9776 &0.633 \\%&8130563
 \hline
\end{tabular}
\end{center}
\end{table}

%\begin{table}[tb]
%\footnotesize
%\begin{center}
%\caption{The effect of kernel size on \emph{HyperPrior} and \emph{Joint} on Kodak, optimized by MS-SSIM with $\lambda=5$.}
%\label{Table.hyper}
%\begin{tabular}{l|l|l|l}
% \hline
% \textbf{Method}   &\textbf{PSNR}& \textbf{MS-SSIM}  & \textbf{Rate}\\%&\textbf{Para}
% \hline
% \emph{HyperPrior-5}        &26.382   &0.9569 &0.170 \\%&5640259
% \hline
% \emph{HyperPrior-9}        &26.266   &0.9591 &0.169 \\%&11188291
% \hline
% \emph{HyperPrior-11}     &26.351   &0.9593 &0.170 \\ %&15151171
% \hline
% \emph{Joint-5}       &26.545   &0.9587 &0.167 \\  %&6623555
% \hline
% \emph{Joint-9}       &26.384   &0.9592 &0.164 \\ %&12171587
% \hline
%\end{tabular}
%\end{center}
%\end{table}

\begin{table}[tb]
\footnotesize
\begin{center}
\caption{The effect of kernel size on \emph{HyperPrior} on Kodak, optimized by MSE with $\lambda=0.015$.}
\label{Table.hyper}
\begin{tabular}{l|l|l|l}
 \hline
 \textbf{Method}   &\textbf{PSNR}& \textbf{MS-SSIM}  & \textbf{Rate}\\%&\textbf{Para}
 \hline
 \emph{HyperPrior-3}        &32.488   &0.9742 &0.543 \\%&5640259
 \hline
 \emph{HyperPrior-5}        &32.976   &0.9757 &0.518 \\%&11188291
 \hline
 \emph{HyperPrior-9}     &33.005   &0.9765 &0.512 \\ %&15151171
 \hline
\end{tabular}
\end{center}
\end{table}

%\begin{table}[tb]
%\footnotesize
%\begin{center}
%\caption{The effect of kernel size with \emph{Joint} architecture, optimized by MS-SSIM with $\lambda=5$.}
%\label{Table.joint}
%\begin{tabular}{|l|l|l|l|l|}
% \hline
% \textbf{Method}  &\textbf{Para}  &\textbf{PSNR}& \textbf{MS-SSIM}  & \textbf{Rate}\\
% \hline
% \emph{Joint-5}    &6623555     &26.545   &0.9587 &0.167 \\
% \hline
% \emph{Joint-9}    &12171587    &26.384   &0.9592 &0.164 \\
% \hline
%\end{tabular}
%\end{center}
%\end{table}

\begin{table}[t]
\footnotesize
\begin{center}
\caption{The effect of kernel size in the auxiliary autoencoder on Kodak, optimized by MS-SSIM with $\lambda=5$.}
\label{Table.hypersub}
\begin{tabular}{l|l|l|l}
 \hline
 \textbf{Method}    &\textbf{PSNR}& \textbf{MS-SSIM}  & \textbf{Rate}\\%&\textbf{Para}
 \hline
 \emph{HyperPrior-9-Aux-5}      &26.266   &0.9591 &0.169 \\%&11188291
 \hline
 \emph{HyperPrior-9-Aux-9}       &26.236   &0.9590 &0.171 \\%&15317059
 \hline
\end{tabular}
\end{center}
\end{table}

\vspace{-1mm}
\subsection{From Shallow Network to Deep Residual Network}

\begin{figure}[tb]
\centering
\subfigure[\emph{ResNet-3$\times$3(3)}]{
\label{Fig.network.1}
\includegraphics[height=150mm]{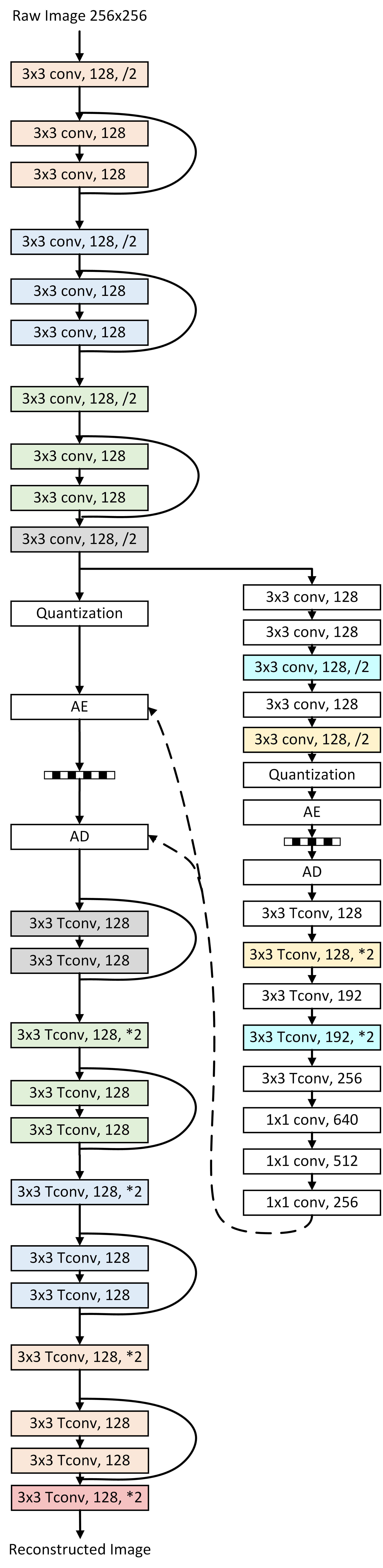}}
\subfigure[\emph{ResNet-3$\times$3(4)}]{
\label{Fig.network.2}
\includegraphics[height=150mm]{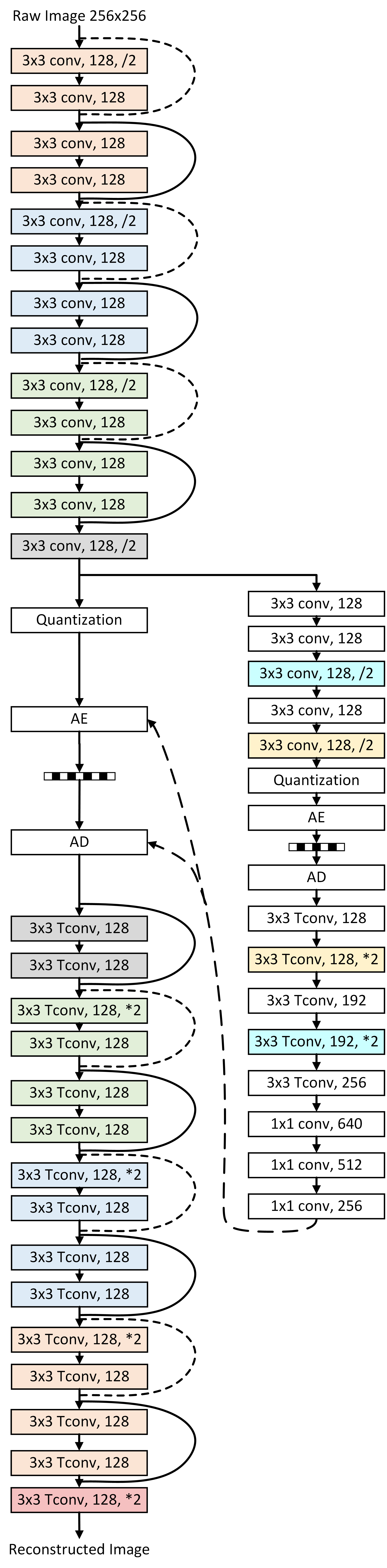}}
\caption{Network structure of proposed deep residual learning, where the solid and dotted lines denote the shortcut connection without and with size change, respectively.}
\label{fig:network}
\end{figure}

\begin{figure}[h]
\centering
\subfigure[Without size change]{
\label{Fig.sub.1}
\includegraphics[height=33mm]{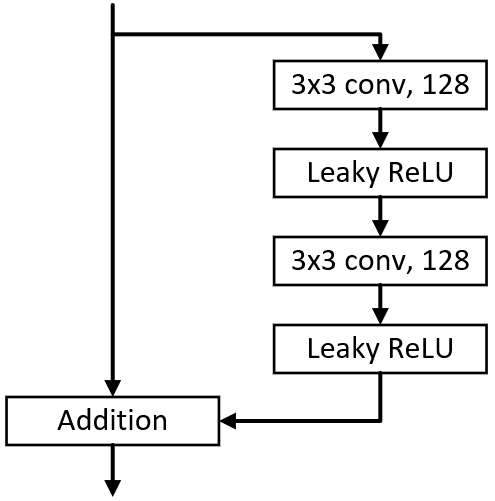}}
\subfigure[With size change]{
\label{Fig.sub.2}
\includegraphics[height=33mm]{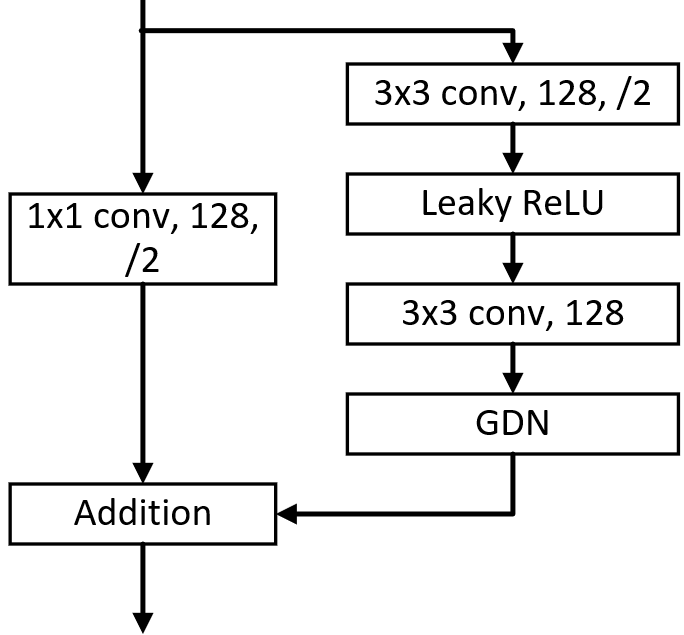}}
\caption{The network structure of one residual unit.}
\label{fig:shortcut}
\end{figure}

With respect to the receptive field, the stack of four $3\times 3$ kernels capture the same receptive field as one $9\times 9$ kernel with fewer parameters. We have tried to replace one large kernel with several $3\times3$ filters, however, experiment shows the stack of $3\times3$ kernels cannot converge. Motivated from~\cite{IEEEexample:ResNet}, we add the shortcut connection for neighboring $3\times3$ kernels. Our proposed deep residual network for image compression is shown in Fig.~\ref{fig:network}. Fig.~\ref{Fig.network.1} is denoted as 3$\times$3(3), where the stack of three $3\times3$ kernels reaches the same receptive field as 7$\times$7. The architecture of Fig.~\ref{Fig.network.2} is \emph{ResNet-3$\times$3(4)}, where the stack of four $3\times3$ kernels reaches the same receptive field as 9$\times$9. As for the activation functions, to prevent more parameters overhead, we only use GDN/IGDN~\cite{IEEEexample:Balle} for one time in each residual unit when the output size changes. For other convolutional layers, we use parameter-free Leaky ReLU as activation function to add the non-linearity in the networks. The shortcut projection is shown in Fig.~\ref{fig:shortcut}. As shown in Table~\ref{Table.subpixel}, \emph{ResNet-3$\times$3(4)} is better than \emph{ResNet-3$\times$3(3)} and \emph{Hyperprior-9}.

\vspace{-1mm}

\subsection{Upsampling Operations at Decoder Side}

The encoder-decoder pipeline is a symmetric architecture. The down-sampling operations at the encoder side are intuitively implemented using convolution filters with stride, however, up-sampling operations at the decoder side have various ways, including bicubic interpolation~\cite{IEEEexample:bicubicSR}, transposed convolution~\cite{IEEEexample:fastSR}, sub-pixel convolution\cite{IEEEexample:SubPixel}. Typically, almost all the previous works use the transposed convolution (\emph{TConv}), except for the work~\cite{IEEEexample:Theis} use sub-pixel convolution at the decoder side. Considering for fast end-to-end learning, we exclude bicubic interpolation and compare two popular up-sampling operations, i.e. TConv and Sub-pixel Conv. For sub-pixel conv, we increase the number of channels by 4 times and then use \emph{tf.depth\_to\_space} function in \emph{Tensorflow}. Results in Table.~\ref{Table.subpixel} show sub-pixel convolution filters bring some improvement on PSNR and MS-SSIM than transposed convolution filters.

%\begin{figure}[tb]
%\centering
%\subfigure[TConv.]{
%\label{Fig.sub.1}
%\includegraphics[height=30mm]{Upsample1.PNG}}
%\hspace{0.4in}
%\subfigure[Subpixel]{
%\label{Fig.sub.2}
%\includegraphics[height=30mm]{Upsample2.PNG}}
%%\subfigure[Interpolation.]{
%%\label{Fig.sub.2}
%%\includegraphics[height=33mm]{Upsample3.PNG}}
%\caption{Different up-sampling operations.}
%\label{fig:upsampling}
%\end{figure}

%\begin{figure}[h]
%	\centerline{\psfig{figure=Subpixel.PNG,width=60.0mm} }
%	\caption{The effect of up-sampling operations.}
%	\label{fig:subpixel}
%\end{figure}

%\begin{figure}[tb]
%\centering
%\subfigure[PSNR vs Rate]{
%\label{Fig.sub.1}
%\includegraphics[width=40mm]{Subpixel_PSNR.PNG}}
%\subfigure[MS-SSIM vs. Rate]{
%\label{Fig.sub.2}
%\includegraphics[width=40mm]{Subpixel_SSIM.PNG}}
%\caption{The effect of up-sampling operations on Kodak.}
%\label{fig:subpixel}
%\end{figure}

%\begin{table}[tb]
%\footnotesize
%\begin{center}
%\caption{Comparison of different residual networks, optimized by MS-SSIM with $\lambda=5$.}
%\label{Table.residual}
%\begin{tabular}{l|l|l|l}
% \hline
% \textbf{Method}   &\textbf{PSNR}  &\textbf{MS-SSIM}  &\textbf{Rate}\\%&\textbf{Para}
% \hline
% \emph{ResNet-3$\times$3(3)}          & 26.378	&0.9605	&0.1704\\ %&6684931
% \hline
% \emph{ResNet-3$\times$3(4)}       & 26.457	&0.9611	&0.1693\\ %&8172172
% \hline
%\end{tabular}
%\end{center}
%\end{table}

\begin{table}[tb]
\footnotesize
\begin{center}
\caption{Comparison of residual networks and upsampling operations on Kodak, optimized by MS-SSIM with $\lambda=5$.}
\label{Table.subpixel}
\begin{tabular}{l|l|l|l}
 \hline
 \textbf{Method}   &\textbf{PSNR}  &\textbf{MS-SSIM}  &\textbf{Rate}\\%&\textbf{Para}
 \hline
 \emph{Hyperprior-9}          & 26.266	&0.9591	&0.1690\\
 \hline
 \emph{ResNet-3$\times$3(3)}          & 26.378	&0.9605	&0.1704\\
 \hline
 \emph{ResNet-3$\times$3(4)-TConv}          & 26.457	&0.9611	&0.1693\\ %&6684931
 \hline
 \emph{ResNet-3$\times$3(4)-SubPixel}       &26.498	&0.9622	 &0.1700\\ %&8172172
 \hline
\end{tabular}
\end{center}
\end{table}

%\begin{table}[tb]
%\begin{center}
%\caption{Model size for different upsampling operations.}
%\label{Table.subpixel}
%\begin{tabular}{|l|l|l|}
% \hline
% \textbf{Method}  &\textbf{Para}  &\textbf{FLOPs}\\
% \hline
% \emph{ResNet-3$\times$3(4)-TConv}       &6684931     &2.43$\times10^{10}$ \\
% \hline
% \emph{ResNet-3$\times$3(4)-SubPixel}    &8172172    &2.50$\times10^{10}$ \\
% \hline
%\end{tabular}
%\end{center}
%\end{table}

\begin{table}[tb]
\footnotesize
\begin{center}
\caption{The effect of wide bottleneck on Kodak dataset.}
\label{Table.bottleneck}
\begin{tabular}{l|l|l|l}
 \hline
 \textbf{Method}    &\textbf{PSNR}& \textbf{MS-SSIM}  & \textbf{Rate}\\%&\textbf{Para}
 \hline
 \emph{ResNet-3x3(4)-Bottleneck128}      &26.498	&0.9622	&0.1700 \\%&8172172
 \hline
 \emph{ResNet-3x3(4)-Bottleneck192}     &26.317	&0.9619	&0.1667\\ %&11627916
 \hline
\end{tabular}
\end{center}
\end{table}

\begin{table}[tb]
\footnotesize
\begin{center}
\caption{Rate control on CLIC validation dataset~\cite{IEEEexample:CLICdata}.}
\label{Table.rate}
\begin{tabular}{l|m{4pt}|m{20pt}|m{34.5pt}|m{20pt}}
 \hline
 \textbf{Method}  &$\boldsymbol{\lambda}$ &\textbf{PSNR}& \textbf{MS-SSIM}  & \textbf{Rate}\\
 \hline
 \emph{ResNet-3x3(4)-Bottleneck192}  &5   &29.708	&0.9697	&0.1369 \\
 \hline
 \emph{ResNet-3x3(4)-Bottleneck192}  &10  &30.710	&0.9765	&0.1816\\
 \hline
\end{tabular}
\end{center}
\end{table}

\begin{table*}[tb]
\footnotesize
\begin{center}
\caption{Results on CLIC validation dataset~\cite{IEEEexample:CLICdata}.}
\label{Table.results}
\begin{tabular}{l|l|l|l|l}
 \hline
 \textbf{Entry}  &\textbf{Description} &\textbf{PSNR}& \textbf{MS-SSIM}  & \textbf{Rate}  \\
 \hline
 \emph{Kattolab}  &\emph{HyperPrior-9}   &28.902	&0.9674	&0.134    \\
 \hline
 \emph{Kattolab}  &\emph{HyperPrior-9} + \emph{Rate Control}   &29.102	&0.9701	  &0.150 \\
 \hline
 \emph{Kattolab}  &\emph{ResNet-3$\times$3(4)-TConv} + \emph{Rate Control}  &29.315	&0.9716	&0.150 \\
 \hline
 \emph{Kattolabv2}  &\emph{ResNet-3$\times$3(4)-SubPixel}+ \emph{Rate Control}  &29.300	&0.9720	&0.150\\
 \hline
 \emph{KattolabSSIM}  &\emph{ResNet-3$\times$3(4)-SubPixel} + \emph{Wide Bottleneck} + \emph{Rate Control} &29.211	&0.9724	&0.150\\
 \hline
\end{tabular}
\end{center}
\end{table*}

\section{Implementation Details}

%\subsection{Training Details}

%We optimized the networks using the distortion metrics MS-SSIM with different values of $\lambda$. Therefore, $\lambda$ is set as $[5, 10, 64, 128]$ to train a total of four models at different bit rates. For each model, average bits per pixel (BPP), and the distortion PSNR and MS-SSIM computed over RGB channels are measured across all the test images.

%The total number of training patches are 13830.
For training, we use $256\times256$ patches cropped from ILSVRC validation dataset (ImageNet~\cite{IEEEexample:ImageNet}). Batch size is 8, and up to 2M iterations are conducted to reach stable results. The model was optimized using Adam~\cite{IEEEexample:adam}, and the learning rate was maintained at a fixed value of $1\times10^{-4}$ and was reduced to $1\times10^{-5}$ for the last 80K iterations.

We also use two strategies for CLIC2019. One is \emph{\textbf{Wide Bottleneck}}. More filters can increase the model capacity. Regarding that increasing the number of filters for large feature maps will significantly increase FLOPs, we only increase the number of filters in the last layer of encoder from 128 to 192, so that FLOPs are only increased a little, from $2.50\times10^{10}$ to $2.56\times10^{10}$. Results are compared in Table.~\ref{Table.bottleneck}. \emph{Bottleneck192} reduces the bitrate a lot, but also degrades quality compared to \emph{Bottleneck128}.

The other is \emph{\textbf{Rate Control}}. For the low-rate track, 0.15bpp is the hard threshold. We train two models at different bit rates by adjusting $\lambda$, where the averaged rate with $\lambda=5$ is less than 0.15bpp for the validation dataset, and the averaged rate with $\lambda=8$ is larger than 0.15bpp. Results are shown in Table.~\ref{Table.rate}. Then we can encode all the test images twice and select adaptive to push the rate to 0.15bpp with the maximized MS-SSIM. One bit should be added into the bitstream to specify which model is used for decoding, which will not increase the complexity of the decoder.

\vspace{-2mm}
\section{Result Analysis}
\vspace{-1mm}

The compression results of our approaches on CLIC validation dataset are summarized in Table~\ref{Table.results}.

%Besides, the 0.15bpp is the hard threshold, so we present a very simple rate allocation strategy to push the limit to 0.15bpp.

Although deep residual network brings the coding gain, the model size grows significantly. In this section, we will analyze the number of parameters and the model complexity with respect to floating point operations per second (FLOPs) for all kinds of architectures. Specifically, take the architecture \emph{HyperPrior-9} as an example, the layer-wise model size analysis is illustrated in Table~\ref{Table.model}. The number of parameters and FLOPs are calculated by
\begin{equation}
\begin{aligned}
&\text{Para} = (h \times w \times C_{in} + 1)\times C_{out}\\
&\text{FLOPs} = \text{Para} \times H' \times W'
\end{aligned}
\end{equation}
where $h\times w$ is the kernel size, $H' \times W'$ is the output size. $C_{in}$ and $C_{out}$ are the number of channels before or after one operation. If no bias is applied, the $+1$ are removed, such as \emph{conv4}. Quantization and leaky-ReLU are parameter-free. GDN~\cite{IEEEexample:Balle3} only run across different channels, but not across different spatial positions, the number of parameters of GDN is only $(C_{in}+1)\times C_{out}$. FLOPs of the total GDN and inverse GDN calculation is only 7.10$\times10^{8}$. This paper mainly focus on the backbone of convolutional layers, so we omit the FLOPs of GDN, inverse GDN and factorized prior. The comparison is listed in Table~\ref{Table.flops}, where the last column is relative value of FLOPs using \emph{Baseline-5}~\cite{IEEEexample:Balle} as a baseline model. \emph{ResNet-3$\times$3(4)} also denotes \emph{ResNet-3$\times$3(4)-TConv}. Our models achieve better coding performance with low complexity.
%The model size of \emph{ResNet-3x3(4)} is much less than \emph{HyperPrior-9}.

\begin{table}[tb]
\footnotesize
\begin{center}
\caption{The model size analysis of \emph{HyperPrior-9}.}
\label{Table.model}
\begin{tabular}{m{25pt}m{2.5pt}m{2.5pt}m{3.5pt}m{3.5pt}m{3.5pt}m{3.5pt}m{24pt}m{30pt}}%|l|l|l|l|l|l|l|l|l|
 \hline
 \textbf{Layer}  & \multicolumn{2}{c}{\textbf{Kernel}} & \multicolumn{2}{c}{\textbf{Channel}} &\multicolumn{2}{c}{\textbf{Output}} & \textbf{Para}  & \textbf{FLOPs}\\
 \cline{2-9}
     &h    &w   &$C_{in}$ &$C_{out}$  &$H'$   &$W'$  &  &  \\
 \hline
 \emph{conv1}    &9    &9   &3 &128 &128 &128   &31232  &5.12$\times10^{8}$\\
 \hline
 \emph{conv2}    &9    &9   &128 &128 &64 &64   &1327232  &5.44$\times10^{9}$\\
 \hline
 \emph{conv3}    &9    &9   &128 &128 &32 &32   &1327232  &1.36$\times10^{9}$\\
 \hline
 \emph{conv4}    &9    &9   &128 &128 &16 &16   &1327104  &3.40$\times10^{8}$\\
 \hline
 \emph{GDN/IGDN} &&&&&& &99072 &- \\
 \hline
 \emph{Hconv1} &3    &3   &128 &128 &16 &16   &147584  &3.78$\times10^{7}$\\
 \hline
 \emph{Hconv2} &5    &5   &128 &128 &8 &8   &409728    &2.62$\times10^{7}$\\
 \hline
 \emph{Hconv3} &5    &5   &128 &128 &4 &4   &409728    &6.56$\times10^{6}$\\
 \hline
 \emph{FactorizedPrior} &&&&&& &5888 &- \\
 \hline
 \emph{HTconv1} &5    &5   &128 &128 &8 &8   &409728    &2.62$\times10^{7}$\\
 \hline
 \emph{HTconv2} &5    &5   &128 &192 &16 &16   &614592  &1.57$\times10^{8}$\\
 \hline
 \emph{HTconv3} &3    &3   &192 &256 &16 &16   &442624  &1.13$\times10^{8}$\\
 \hline
 \emph{layer1} &1    &1   &256 &640 &16 &16   &164480  &4.21$\times10^{7}$\\
 \hline
 \emph{layer2} &1    &1   &640 &512 &16 &16   &328192  &8.40$\times10^{7}$\\
 \hline
 \emph{layer3} &1    &1   &512 &256 &16 &16   &131072  &3.36$\times10^{7}$\\
 \hline
 \emph{Tconv1}         &9    &9   &128 &128 &32 &32   &1327232  &1.36$\times10^{9}$\\
 \hline
 \emph{Tconv2}         &9    &9   &128 &128 &64 &64   &1327232  &5.44$\times10^{9}$\\
 \hline
 \emph{Tconv3}         &9    &9   &128 &128 &128 &128   &1327232  &2.17$\times10^{10}$\\
 \hline
 \emph{Tconv4}         &9    &9   &128 &128 &256 &256   &31107  &2.04$\times10^{9}$\\
 \hline
 \textbf{Total}    &&&&&& &\textbf{11188291} &\textbf{3.88}$\boldsymbol{\times10^{10}}$ \\
 \hline
\end{tabular}
\end{center}
\end{table}

\begin{table}[tb]
\footnotesize
\begin{center}
\caption{The model complexity of different architectures.}
\label{Table.flops}
\begin{tabular}{m{85pt}|m{30pt}|l|c}
 \hline
 \textbf{Method}  &\textbf{Para}  &\textbf{FLOPs}  &\textbf{Relative}\\
 \hline
 \emph{Baseline-3}    &997379     &4.25$\times10^{9}$   &0.36 \\
 \hline
 \emph{Baseline-5}    &2582531    &1.18$\times10^{10}$  & 1.00\\
 \hline
 \emph{Baseline-9}    &8130563    &3.82$\times10^{10}$   &3.24 \\
 \hline
 \emph{HyperPrior-3}    &4055107    &4.78$\times10^{9}$  &0.40\\
 \hline
 \emph{HyperPrior-5}    &5640259     &1.23$\times10^{10}$  &1.04  \\
 \hline
 \emph{HyperPrior-9}    &11188291    &3.88$\times10^{10}$  &3.28\\
 \hline
 \emph{ResNet-3$\times$3(3)}   &5716355   &1.75$\times10^{10}$  &1.48 \\
 \hline
 \emph{ResNet-3$\times$3(4)}   &6684931   &2.43$\times10^{10}$  &2.06 \\
 %\hline
 %\emph{ResNet-3$\times$3(4)-TConv}       &6684931     &2.43$\times10^{10}$ \\
 \hline
 \emph{ResNet-3$\times$3(4)-SubPixel}    &8172172    &2.50$\times10^{10}$  &2.12\\
 \hline
 \emph{ResNet-3$\times$3(4)-SubPixel-Bottleneck192}    &11627916    &2.56$\times10^{10}$  &2.17\\
 \hline
\end{tabular}
\end{center}
\end{table}

\vspace{-2mm}

\section{Conclusion}

In this paper, we have described the proposed deep residual learning and sub-pixel convolution for image compression. This is the basis of our submitted entries \emph{Kattolab}, \emph{Kattolabv2} and \emph{KattolabSSIM}. Results have shown our approaches achieve 0.972 in MS-SSIM at the rate of 0.15bpp with moderate complexity during the validation phase.

%
%{\small
%\bibliographystyle{ieee_fullname}
%\bibliography{egbib}
%}

\end{document}